\documentclass[english,prd, twocolumn,showpacs,showkeywords,showpreprint]{revtex4}
\usepackage[T1]{fontenc}
\usepackage[latin9]{inputenc}
\usepackage[active]{srcltx}
\usepackage{textcomp}
\usepackage{amsmath}
\usepackage{amssymb}
\usepackage{graphicx}
\usepackage{esint}

\makeatletter
\@ifundefined{textcolor}{}
{%
 \definecolor{BLACK}{gray}{0}
 \definecolor{WHITE}{gray}{1}
 \definecolor{RED}{rgb}{1,0,0}
 \definecolor{GREEN}{rgb}{0,1,0}
 \definecolor{BLUE}{rgb}{0,0,1}
 \definecolor{CYAN}{cmyk}{1,0,0,0}
 \definecolor{MAGENTA}{cmyk}{0,1,0,0}
 \definecolor{YELLOW}{cmyk}{0,0,1,0}
}

\makeatother

\usepackage{babel}
\begin{document}

\title{QCD condensates and holographic Wilson loops for asymptotically AdS
spaces }

\author{R. Carcasses Quevedo$^{a,b}$ }

\email{robert.carcasses-quevedo@ib.edu.ar}

\selectlanguage{english}%

\author{J. L. Goity $^{c,d}$ }

\email{goity@jlab.org}

\selectlanguage{english}%

\author{R. C. Trinchero $^{a,b}$ }

\email{trincher@cab.cnea.gov.ar }

\selectlanguage{english}%

\affiliation{$^{a}$Instituto Balseiro, Centro Atómico Bariloche, 8400 San Carlos
de Bariloche, Argentina.}

\affiliation{$^{b}$CONICET, Rivadavia 1917, 1033 Buenos Aires, Argentina..\\
 $^{c}$Department of Physics, Hampton University, Hampton, VA 23668,
USA}

\affiliation{$^{d}$Thomas Jefferson National Accelerator Facility, Newport News,
VA 23606, USA.\\
 }
\begin{abstract}
The minimization of the Nambu-Goto action for a surface whose contour
defines a circular Wilson loop of radius $a$ placed at a finite value
of the coordinate orthogonal to the boundary is considered. This is
done for asymptotically AdS spaces. The condensates of even dimension
$n=2$ through 10 are calculated in terms of the coefficient of $a^{n}$
in the expansion of the on-shell subtracted Nambu-Goto action for
small $a$. The subtraction employed is such that it presents no conflict
with conformal invariance in the AdS case and need not introduce an
additional infrared scale for the case of confining geometries. It
is shown that the UV value of the condensates is universal in the
sense that they only depends on the first coefficients of the difference
with the AdS case.
\end{abstract}



\preprint{\noindent Publication number: JLAB-THY-13-1819}

\maketitle

\section{Introduction}

The relation between large $N$ gauge theories and string theory \cite{'tHooft:1973jz}
together with the AdS/CFT correspondence \cite{Maldacena:1997re,Witten:1998qj,Gubser:1998bc,malda:1999ti}
have opened new insights into strongly interacting gauge theories.
The application of these ideas to QCD has received significant attention
since those breakthroughs. From the phenomenological point of view,
the so called AdS/QCD approach has produced very interesting results
in spite of the strong assumptions involved in its formulation \cite{Gubser:1999pk,Sakai:2004cn,Da_Rold:2005zs,Erlich:2005qh,Polchinski:2000uf,Csaki:2006ji}.
It seems important to further proceed investigating these ideas and
refining the current understanding of a possible QCD gravity dual. 

As is well known the vacuum of pure gauge QCD is the simplest setting
that presents key non-perturbative effects of QCD. In this regard,
the gluon condensate $G_{2}\equiv\frac{g^{2}}{4\pi^{2}}\langle F_{\mu\nu}^{a}F_{a}^{\mu\nu}\rangle$
plays an important role. The existence of a non-vanishing $G_{2}$
was early on identified \cite{Shifman1979385}. It has important manifestations
in hadron phenomenology \cite{Colangelo:2000dp,Shifman1979385}, and
there are indications of its non-vanishing from lattice QCD \cite{DiGiacomo:1981wt,Rakow:2005yn,Banks:1981zf}.
The gluon condensate can be obtained from the vacuum expectation value
of a small Wilson loop. In the holographic approach, such an expectation
value is obtained by minimizing the Nambu-Goto (NG) action for a loop
lying in the boundary space \cite{Maldacena:1998im,Rey:1998ik}. This
is known to work in the strictly AdS case, i.e. for a conformal boundary
field theory. In this work we assume that this procedure also works
in the non-conformal-QCD case provided an adequate 5-dimensional background
metric is chosen. 

The features and results of this work are summarized as follows:
\begin{itemize}
\item The NG action for a circular loop of radius $a$ lying at a given
value of the coordinate orthogonal to the boundary of an asymptotically
AdS space is considered. 
\item The minimization of this action leads to equations of motion, whose
solution is approximated by a power series in $a$.
\item The on-shell NG action is subtracted following the procedure in ref.
\cite{Maldacena:1998im}. More precisely, an extension of this procedure
is proposed for the case under consideration, where the base of the
loop is at a finite value of the radial coordinate and a natural infrared
limit is considered for the case of confining theories.
\item The gluon condensates of even dimension $n=2$ through 10 are obtained
from the coefficients of the expansion in powers of $a$ of the subtracted
on-shell NG action $S_{NG}^{sub}$, the last four ones assuming the
absence of the condensate of dimension $2$.
\item It is shown that the UV value of these condensates is universal in
the sense that for a condensate of a given dimension, its value does
not depend on the value of the warp factor's higher order coefficients.
\end{itemize}
The paper is organized as follows. Section II defines the problem
to be considered, including the NG action for the circular loop and
the asymptotically AdS background metric. Section III deals with the
subtraction of the on-shell NG action. Section IV gives some model
independent results, which clarify the relation between condensates
and the expansion coefficients of the warp factor. Section V deals
with the approximate solution of the equations of motion and the evaluation
of the on-shell NG action as a power series in $a$. Section VI gives
the results for the gluon condensates, showing the above mentioned
universality. Section VII includes some concluding remarks. In addition
four appendices are included.

\section{NG action for a circular loop on an asymptotically $AdS$ space}

The distance to be considered has the following general form,
\begin{eqnarray}
ds^{2} & = & e^{2A(z)}(dz^{2}+\eta_{ij}dx^{i}dx^{j})\nonumber \\
 & = & G_{\mu\nu}dx^{\mu}dx^{\nu}\,\,\,\,\,\,\,\,\,\,\;\;\mu,\nu=1,\cdots,d\,+1\;\quad.
\end{eqnarray}
It is defined by a metric with no dependence on the boundary coordinates
and preserves the boundary space Poincare invariance. This should
be the case if only vacuum properties are considered. The form of
the warp factor $A(z)$ to be considered is,
\begin{equation}
A(z)=-\ln\left(\frac{z}{L}\right)+f(z),\label{eq:warp}
\end{equation}
where $f\left(z\right)$ is a dimensionless function. In this work
$f(z)$ is taken to be a power series in $z$  %
\footnote{Recalling that near the UV boundary the relation between the conformal
coordinate $z$ and Fefferman-Graham \cite{Fefferman-Graham:84} coordinate
$\rho$ is $\rho=z^{2}$, then a polynomial in $z$, as considered
in this work, corresponds to an expression involving integer and half
integer powers of $\rho$. However see section IV where it is shown
that half integers powers of $\rho$ can not appear for a theory describing
QCD.%
}, i.e.,
\begin{equation}
f(z)=\sum_{k=1}\alpha_{k}z^{k}\;.\label{eq:fz}
\end{equation}
 The case $f\left(z\right)=0$ corresponds to the $AdS$ metric. This
deviation from the AdS case could be produced by a bulk gravity theory
including matter fields \cite{hung-myers-smolkin:11}. Possible candidates
for these bulk gravity theories have been considered in \cite{Gursoy:2007cb,Goity:2012yj}. 

The area of a surface embedded in this space is given by the NG action,
\begin{equation}
S_{NG}=\frac{1}{2\pi\alpha'}\int d^{2}\sigma\sqrt{g}\,,\label{eq:action}
\end{equation}
where $g$ is the determinant of the induced metric on the surface,
which is given by,
\[
g_{ab}=G_{\mu\nu}\partial_{a}X^{\mu}\partial_{b}X^{\nu}\;,
\]
where $X^{\mu}$ are the coordinates of the surface embedded in the
ambient $d+1$ dimensional space. The indices $a,b$ refer to coordinates
on the surface. The case to be considered is a circular loop whose
contour lies at a constant value $z_{1}$ of the coordinate $z$ and
in the $i-j$ spatial plane. The coordinates on the surface are then
taken to be $r$ and $\phi$, the polar coordinates. Therefore the
embedding can be described by the following,
\begin{eqnarray}
X^{k} & = & 0\label{eq:embedd}\\
X^{5} & = & z\left(r\right)\\
\, X^{i} & = & r\cos\phi,\;\;\,\, X^{j}=r\sin\phi\,\,\,\,\,\,\,(\forall k\neq i\neq j),
\end{eqnarray}
with the boundary conditions,
\begin{equation}
z(a)=z_{1},\,\,\,\,\,\,\,\,\, z'(0)=0\;,
\end{equation}
which states that the contour of the circular loop of radius $a$
is located at $z_{1}$ and that no cusps are admitted. Replacing the
embedding (\ref{eq:embedd}) in the action (\ref{eq:action}), after
a trivial integration in the angular variable, leads to the following
expression,
\begin{equation}
S_{NG}=\frac{1}{\alpha'}\int_{0}^{a}e^{2A\left(z\right)}r\sqrt{1+z'^{2}}dr\;,\label{eq:actionr}
\end{equation}
where the prime denotes derivative respect to $r$. The minimal surface
is given by the solution of the following equations of motion with
the above mentioned boundary conditions,
\begin{eqnarray}
r\frac{z''\left(r\right)}{1+z'\left(r\right)^{2}}+z'\left(r\right)-2r\frac{dA\left(z\right)}{dz} & = & 0\;.\label{eq:motion}
\end{eqnarray}
For the AdS case $A(z)=-\ln\frac{z}{L}$ the solution is,
\begin{equation}
z(r)=\sqrt{a^{2}+z_{1}^{2}-r^{2}}\;,\label{eq:solads}
\end{equation}
which upon replacing in (\ref{eq:action}) leads to the following
expression for the on-shell NG action,
\begin{equation}
S_{NG_{AdS}}^{o.s.}=\frac{L^{2}}{\alpha'}\left(\sqrt{1+\frac{a^{2}}{z_{1}^{2}}}-1\right).\label{eq:nambu-goto evaluada en ads}
\end{equation}

\section{The subtracted on-shell NG action\label{sec:The-substracted-on-shell}}

Replacing the solution of the previous section in the NG action leads
to a divergent expression when $z_{1}\to0$, i.e. near the UV boundary
of the space. This happens in the AdS case and also when $f(z)\neq0$
. Therefore, a subtraction procedure should be employed. The action
requires regularization, where the most obvious procedure is to choose
$z_{1}\neq0$, and a renormalized action is obtained by implementing
a subtraction, as it is discussed in detail in this section. A procedure
of minimal subtraction, implemented by disregarding the $1/z_{1}$
term in $S_{NG}$ was implemented in \cite{Andreev:2007vn}. 

\begin{figure}
\begin{centering}
\includegraphics[width=8cm]{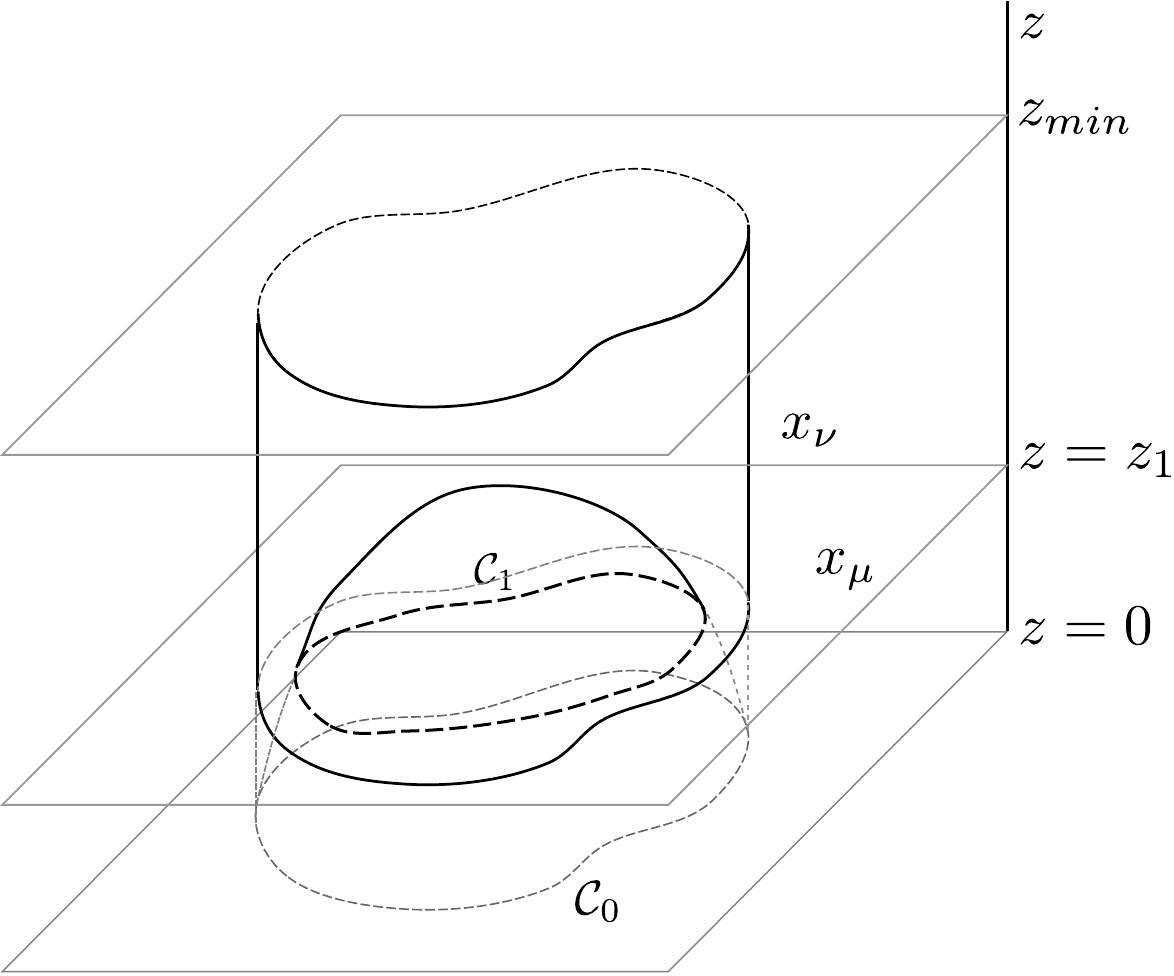}
\par\end{centering}

\caption{Substraction scheme.\label{fig:Substraction-schema.}}
\end{figure}

As shown in \cite{Maldacena:1998im} for the rectangular loop, a physically
motivated procedure is to subtract the contribution of the heavy \textquotedbl{}quark\textquotedbl{}
mass to the action. This contribution corresponds to the area of a
cylinder with axis parallel to $z$, extending from $z=\infty$ to
$z=0$, for the case of the base of the loop located at $z=0$. It
could be thought that in the case considered in this paper, the base
of the loop located at $z_{1}$, the area of a cylinder with axis
parallel to $z$, extending from $z=\infty$ to $z_{1}$, should be
subtracted. However such a procedure should be modified in two aspects,
\begin{itemize}
\item First, in the AdS case, it leads to a loss of conformal invariance,
more precisely the value of $S_{NG}^{sub}$ would depend on the radius
of the loop. Requiring independence of the value of $S_{NG}^{sub}$
on the radius $a$, leads to the following definition of the subtracted
action,
\begin{equation}
S_{NG}^{sub}=S_{NG}-\frac{r_{0}(a,z_{1})}{\alpha'}\int_{z_{1}}^{z_{IR}}dz\, e^{2A(z)}\;,\label{eq:nambu-goto substraida}
\end{equation}
where $z_{IR}$ is an infrared scale whose motivation and definition
is explained below. For the AdS case the function $r_{0}(a,z_{1})$
is fixed by conformal invariance and given by,
\begin{equation}
r_{0}^{AdS}(a,z_{1})=\sqrt{a^{2}+z_{1}^{2}}\;,\label{eq:rads}
\end{equation}
leading to,
\begin{equation}
S_{NG}^{sub,AdS}=-\frac{L^{2}}{\alpha'}\quad.\label{eq:NGadssubs}
\end{equation}
The radius $r_{0}(a,z_{1})$ corresponds to the radius of a loop located
at the boundary whose minimal surface would intersect the plane $z=z_{1}$
with a circle of radius $a$. Therefore the following holds,
\begin{equation}
\lim_{z_{1}\to0}r_{0}(a,z_{1})=a\;.\label{eq:limr0}
\end{equation}
For non-AdS cases the same procedure could be employed, using the
corresponding value $r_{0}^{Non-AdS}(a,z_{1})$. However it should
be noted that in the non-AdS case using the AdS $r_{0}$ given in
(\ref{eq:rads}), also leads to a finite value for the $S_{NG}^{sub}$
 and presents no-conflict with conformal invariance. This last procedure
will be employed below.
\item Second, confining warp factors are such that $e^{A(z)}$ presents
a global minimum for a finite value for this factor\cite{kinar-sonnenschein:1998,Gursoy:2007er}.
Let $z_{m}$ denote the location of this minimum in the coordinate
$z$. In these conditions, integrating $e^{2A(z)}$ between $z_{1}$
and $\infty$ would lead to a divergent result. Introducing a infrared
integration limit $z_{IR}$ as in (\ref{eq:nambu-goto substraida})
eliminates this divergence at the cost of introducing this ad-hoc
infrared cut-off. In this respect the following remarks are important,

\begin{itemize}
\item The result for the coefficients of $a^{n}\;,n>1$ in  $S_{NG}^{sub}$
do not depend on $z_{IR}$. This fact is shown in section \ref{sec:The-Gluon-condensate,},
and is due to property (\ref{eq:limr0}). 
\item On the other hand these confining warp factors already have a natural
infrared scale. This is given by the location $z_{m}$ of the global
minimum. This is a natural candidate to be identified with $z_{IR}$.
In this respect it is worth noting that for $z_{1}<z_{m}$ the minimal
surface could never exceed the value $z_{m}$, otherwise it would
not be minimal %
\footnote{Indeed, suppose there were a minimal surface with boundary at $z_{1}<z_{m}$
that extends to values of $z>z_{m}$, then, since the warp factor
necessarily grows for these values(recall that $z_{m}$ is a minimum),
a surface stopping at $z_{m}$ will have less area than the one originally
supposed to be minimal, which is a contradiction. %
}. In what follows the choice $z_{IR}=z_{m}$ is made. It is emphasized
that other choices are by no means excluded. Different choices produce
different coefficients for the perimeter in $S_{NG}^{sub}$.
\end{itemize}
\end{itemize}
It is noted that this subtraction is non-vanishing even if the loop
is located at a finite value of the coordinate $z$. Figure \ref{fig:Substraction-schema.}
illustrates the proposed subtraction procedure, for the case of a
minimal surface bounded by a closed contour $\mathcal{C}_{1}$ .

\subsection{Convergence of the subtracted NG action in the UV limit\label{sub:convergence proof}}

The considered warp factors diverge in the UV, the leading singularity
is,

\begin{equation}
A\left(z\right)\sim-\ln\left(\frac{z}{L}\right)\Rightarrow A'\left(z\right)\sim-\frac{1}{z}\;,\label{eq:az0}
\end{equation}
this makes the integrand appearing in the NG action diverge at $z=0$
. 

In order to analyze the behavior of the solution near the boundary,
the approach in \cite{Graham-Witten:99} is employed. The NG action
is written in terms of $r$ as a function of $z$, $r\left(z\right)$,
this leads to,

\begin{equation}
S_{NG}=\frac{1}{\alpha'}\int_{z_{1}}^{z_{0}}e^{2A\left(z\right)}r\left(z\right)\sqrt{1+r'\left(z\right)^{2}}dz\;,
\end{equation}
where $r\left(0\right)=r_{0}$ and $r\left(z_{1}\right)=a$. The equation
of motion is,
\begin{equation}
\left(-1+2rA'\left(z\right)r'\right)\left(1+r'^{2}\right)+rr''=0\;.\label{eq:em r en funcion de z-1}
\end{equation}
Near the boundary (\ref{eq:em r en funcion de z-1}) implies,
\begin{equation}
-\underset{z\rightarrow0}{\lim}2r\frac{1}{z}r'\left(1+r'^{2}\right)+\underset{z\rightarrow0}{\lim}rr''-\underset{z\rightarrow0}{\lim}\left(1+r'^{2}\right)=0\;.\label{eq:em en el limite z->0-1}
\end{equation}

This last equation shows that if $\lim_{z_{1}\to0}r''\left(z\right)$
is assumed to be finite, then $\lim_{z_{1}\to0}r'\left(z\right)$
can not be infinite since in that case it will be impossible to cancel
the terms containing $r'$. Furthermore the cancellation of the terms
involving $r'\left(z\right)$, require that $\lim_{z_{1}\to0}r'\left(z\right)=0$
as $z^{1+\epsilon}\;,\epsilon>0$. In addition the cancellation of
the constant term in (\ref{eq:em en el limite z->0-1}) requires $\epsilon=0$. 

On the contrary, if $\lim_{z\rightarrow0}r''\left(z\right)$ is assumed
to be infinite, then $\lim_{z_{1}\to0}r'\left(z\right)\to\infty$,
which can be proved by integrating the former, and again it is not
possible to cancel all the divergent terms due to their different
degrees of divergence. Therefore,
\begin{eqnarray}
\underset{z\to0}{\lim}r'\left(z\right) & = & 0\\
r'\left(z\right) & =- & \frac{z}{a}+...\;\left(z\ll a\right)\quad.
\end{eqnarray}

Next, this asymptotics is plugged in the $S_{NG}^{sub}$ (\ref{eq:nambu-goto substraida}).
In this respect it is convenient to rewrite it in the form,

\begin{eqnarray*}
S_{NG}^{sub} & = & \frac{1}{\alpha'}\int_{z_{1}}^{z_{0}}e^{2A\left(z\right)}\left(r\left(z\right)\sqrt{1+r'\left(z\right)^{2}}-r_{0}(a,z_{1})\right)\\
 &  & -\frac{r_{0}(a,z_{1})}{\alpha'}\int_{z_{0}}^{z_{m}}dz\, e^{2A(z)}\quad.
\end{eqnarray*}
For the considered cases of $A\left(z\right)$, the second term is
convergent because the integrand has no poles in the finite integration
interval. The first term is also finite, indeed:

\begin{eqnarray*}
\underset{z\to0}{\lim}e^{2A\left(z\right)}\left(r\left(z\right)\sqrt{1+r'\left(z\right)^{2}}-a\right) & =\\
\underset{z\to0}{\lim}\frac{1}{z^{2}}\left(a\left(1+\frac{1}{2}c_{0}^{2}z^{2}+\dots\right)-a\right) & =\\
a\underset{z\to0}{\lim}\frac{1}{z^{2}}\left(\frac{1}{2}c_{0}^{2}z^{2}+\dots\right) & = & 0\quad.
\end{eqnarray*}
Thus, the integrand is finite everywhere inside the finite integration
region and therefore the integral is finite. Furthermore this last
equation shows that the divergent term in the UV of $S_{NG}$ is proportional
to the perimeter of the loop, which is consistent with the fact that
also the subtraction is proportional to the perimeter of the loop
in the UV, as shown by (\ref{eq:limr0}).

\section{Model Independent Results}

In this section some results that follow from the general setting
described in the previous sections are considered. No approximation
is involved in the derivation of these properties.

\subsection{In QCD $f(z)$ is even\label{sub:In-QCD-}}

It is recalled that $f(z)$ is the function appearing in the warp
factor (\ref{eq:warp}). The title of this subsection means the following.
The basic hypothesis underlying this work is that the vacuum expectation
value of the Wilson loop in QCD is given by  $S_{NG}^{sub}$ . It
will be shown below that under this assumption, the fact that there
are no odd-dimensional condensates %
\footnote{By definition a condensate of dimension $n$ is the coefficient of
$a^{n}$ in the expansion of $\frac{\alpha'}{L^{2}}S_{NG}[z](a)$
in powers of a.%
} in QCD implies that $f(z)=f(-z)$. The proof of this assertion is
based on the following intermediate result.

\emph{If the expansion of $\frac{\alpha'}{L^{2}}S_{NG}[z](a)$ as
a power series in $a$ only involves even powers of $a$ then,}
\begin{equation}
f(z)-f(-z)=const.\label{eq:no-odd}
\end{equation}

Proof. Denoting by $S_{NG}[z](a)$ the NG action with parameter $a$,
the hypothesis is,
\[
S_{NG}[z](a)=S_{NG}[z](-a)\;.
\]
Noting that the change $a\to-a$ is, at the level of the NG action,
the same as changing $z\to-z$ implies, 
\begin{eqnarray*}
S_{NG}[z](-a) & = & S_{NG}[-z](a)\\
 & \Downarrow & (\ref{eq:no-odd})\\
S_{NG}[z](a) & = & S_{NG}[-z](a)\;.
\end{eqnarray*}
Due to this last equality if $z(r)$ extremizes the NG action so does
$-z(r)$. Therefore $-z(r)$ must also be a solution of the equation
of motion. The equation of motion for $z(r)$ is,
\[
r\frac{z''\left(r\right)}{1+z'\left(r\right)^{2}}+z'\left(r\right)-2r(-\frac{1}{z}+\frac{df}{dz}\left(z\right))=0
\]
the equation of motion for $-z(r)$ is,
\[
-\left(r\frac{z''\left(r\right)}{1+z'\left(r\right)^{2}}+z'\left(r\right)-2r(-\frac{1}{z}-\frac{df}{dz}\left(-z\right)\right)=0
\]
summing up these two equations leads to,
\begin{eqnarray*}
\frac{df}{dz}\left(-z\right)- & \frac{df}{dz}\left(z\right)= & 0\Rightarrow\frac{d}{dz}[f(-z)-f(z)]=0\\
 & \Downarrow\\
f(-z)-f(z) & = & \mathrm{const}.
\end{eqnarray*}
as claimed.

Now, the only solution to (\ref{eq:no-odd}) for arbitrary $z$ is,
\[
f_{odd}(z)=f(z)-f(-z)=0\;,
\]
which shows that only even functions $f(z)$ are relevant for QCD.
In particular for the warp factors considered in the present work,
the above general result implies that if the only non-vanishing condensates
are even dimensional, the coefficients $\alpha_{n}$ must vanish if
$n=\mathrm{odd.}$

\subsection{Condensate of dimension $n>1$ is independent of $\alpha_{m}$ for
$m>n$ and $z_{1}\to0$ \label{sub:UV-universality}}

First it is noted that $\frac{\alpha'}{L^{2}}S_{NG}[z](a,\alpha)$
is dimensionless and that $\alpha_{n}$ has dimension of length to
the $-n$. Therefore if $\alpha_{m}$ would contribute to the condensate
of dimension $n<m$ then inverse powers of $\alpha_{k}$ should appear
for some $k>n$. Therefore in that case $\frac{\alpha'}{L^{2}}S_{NG}[z](a,\alpha)$
would diverge when $\alpha_{k}\to0$. However the integrand in $\frac{\alpha'}{L^{2}}S_{NG}[z](a,\alpha)$
is well defined when any or all of the $\alpha$'s vanish. Indeed
the only divergence in $\frac{\alpha'}{L^{2}}S_{NG}[z](a,\alpha)$
is proportional to $a$, and appears when all the $\alpha_{n}$ vanish,
but something proportional to $a$ does not contribute to the condensates
with $n>1$. Therefore only positive powers of the $\alpha$'s can
appear and the result follows from dimensional reasons. It should
be noted that this result holds for $z_{1}\to0$, otherwise since
$z_{1}$ has dimensions of length all the dimensional arguments made
above are not valid. In conclusion, the general expression for the
expansion in powers of $a$ of the NG action is, 
\begin{eqnarray*}
\frac{\alpha'}{L^{2}}S_{NG}[z](a,\alpha) & = & s^{(0)}+s^{(2)}\alpha_{2}a^{2}+(s_{2}^{(4)}\alpha_{2}^{2}+s_{4}^{(4)}\alpha_{4})a^{4}\\
 &  & +(s_{2}^{(6)}\alpha_{2}^{3}+s_{2,4}^{(6)}\alpha_{2}\alpha_{4}+s_{6}^{(6)}\alpha_{6})a^{6}+\cdots\;,
\end{eqnarray*}
where the coefficients $s^{(n)}$ are dimensionless.

\section{On-shell NG action expanded in powers of the radius $a$}

\subsection{Condensates of dimension $2$ and $4$ }

The approach employed in this section is basically the same as in
\cite{Andreev:2007vn}. That is, expand the solution of the equations
of motion as a power series in $a^{2}$, replace in the Lagrangian,
expand it in powers of $a^{2}$ and then integrate. However they differ
in some aspects. An important difference is that in this work more
general curved backgrounds are considered. More precisely, the warp
factors given in (\ref{eq:warp})-(\ref{eq:fz}) are considered for
$n=1$ and for both $\alpha_{2}$ and $\alpha_{4}$ non-vanishing.
The consideration of $\alpha_{4}\neq0$ is particularly relevant from
the phenomenological point of view. This is so because $\alpha_{4}\neq0$
allows for a non-vanishing gluon condensate of dimension $4$ without
having at the same time one of dimension 2 which is not allowed in
QCD %
\footnote{In this assertion the effect of renormalons is neglected. This assumption
is supported by the results in \cite{Rakow:2005yn,Ilgenfritz:2012nca}.\label{fn:In-this-assertion}%
}. The other difference concerns the subtraction procedure which in
this work is done as described in section \ref{sec:The-substracted-on-shell}.
According to this procedure the NG action should be calculated for
a loop lying at a value $z_{1}$ of the coordinate orthogonal to the
boundary. In this respect it is convenient to define the variable,
\[
t=\sqrt{1+w_{1}^{2}-\rho^{2}}\;\;,w_{1}=\frac{z_{1}}{a}\,,\rho=\frac{r}{a}\;.
\]
In this variable the AdS solution (\ref{eq:solads}) is written as,
\[
w(t)=t\;\;,w=\frac{z}{a}\;.
\]
In terms of the variable $\psi(t)=w^{2}(t)$ the NG action is given
by:

\begin{equation}
S_{NG}=\frac{L^{2}}{\alpha'}\int_{0}^{1}\frac{e^{2(a^{2}\alpha_{2}\psi+a^{4}\alpha_{4}\psi^{2})}t\sqrt{4+\frac{\left(1+w_{1}^{2}-t^{2}\right)\psi'(t)^{2}}{t^{2}\psi\left(t\right)}}}{2\psi\left(t\right)}dt\;.\label{eq:NG en funcion de t}
\end{equation}
The equation of motion for this action reads:

\begin{eqnarray}
64a^{4}t^{3}\alpha_{4}\psi(t)^{3}-\left(1-t^{2}+w_{1}^{2}\right)\left(2t-\psi'(t)\right)\psi'(t)^{2}+\nonumber \\
+16\psi(t)^{2}\left(2a^{2}t^{3}\alpha_{2}+a^{4}t\left(1-t^{2}+w_{1}^{2}\right)\alpha_{4}\psi'(t)^{2}\right)-\nonumber \\
-4\psi(t)\left(4t^{3}-\left(1+t^{2}+w_{1}^{2}\right)\psi'(t)\right)-\nonumber \\
-4\psi(t)\left[-2a^{2}t\left(1-t^{2}+w_{1}^{2}\right)\alpha_{2}\psi'(t)^{2}\right.\label{eq:motionpsi}\\
\left.+t\left(1-t^{2}+w_{1}^{2}\right)\psi''(t)\right] & = & 0\;.\nonumber 
\end{eqnarray}
As explained earlier, the boundary conditions to be required are the
following ones
\[
\psi(w_{1})=w_{1}^{2}\;,\quad\psi'(\sqrt{1+w_{1}^{2}})=\mathrm{finite\;,}
\]
which correspond to the loop located in the plane $z=z_{1}$ and the
surface with no cusp at $r=0$. 

Next a power series expansion ansatz for the solution is considered
,
\begin{equation}
\psi\left(t\right)=\sum_{i=0}\psi_{i}\left(t\right)a^{2i}\label{eq:solpsi}
\end{equation}
Replacing in (\ref{eq:motionpsi}) and requiring the vanishing of
the coefficient in front of $a^{2i}$, for $i=0$ this leads to,
\begin{eqnarray*}
\left(-1+t^{2}-w_{1}^{2}\right)\left(2t-\psi_{0}'(t)\right)\psi_{0}'(t){}^{2}\\
+4\psi_{0}(t)\left[\left(\left(1+t^{2}+w_{1}^{2}\right)\psi_{0}'(t)\right)\right.\\
\left.+t\left(-4t^{2}+\left(-1+t^{2}-w_{1}^{2}\right)\psi_{0}''(t)\right)\right] & = & 0
\end{eqnarray*}
whose solution is the AdS one $\psi_{0}(t)=t^{2}$. For $i=1$,
\begin{eqnarray*}
2\left(1+w_{1}^{2}\right)\left(4t^{3}\alpha_{2}+\psi_{1}'(t)\right) & =\\
+t\left(-1+t^{2}-w_{1}^{2}\right)\psi_{1}''(t) & = & 0
\end{eqnarray*}
whose solution up to order $\mathcal{O}(w_{1}^{2})$ is,
\begin{eqnarray*}
\psi_{1}(t) & = & -\frac{1}{1+t}4\left\{ t\left(-2-t+t^{2}+(-4+(-2+t)t)\, w_{1}^{2}\right)\right.\\
 &  & +2(1+t)\left(1+2w_{1}^{2}\right)\mathrm{arctanh}(t)\\
 &  & \left.+(1+t)\left(1+2w_{1}^{2}\right)\log(1-t^{2})\right\} \alpha_{2}\;.
\end{eqnarray*}
In a similar fashion the equation and its solution for $\psi_{2}(t)$
are obtained.

Next the NG action expansion in powers of $a$ is computed. Replacing
the solution (\ref{eq:solpsi}) in the integrand of (\ref{eq:NG en funcion de t}),
expanding in powers of $a$ and $w_{1}$ and integrating leads to,

\begin{eqnarray}
S_{NG} & = & \frac{L^{2}}{\alpha'}\left(\frac{\sqrt{1+w_{1}^{2}}}{w_{1}}-1\right.\nonumber \\
 &  & +a^{2}\left[\frac{10}{3}-w_{1}^{2}(\frac{8}{3}-7+\log(16))\right]\alpha_{2}\nonumber \\
 &  & +a^{4}\left\{ \left[\frac{14}{9}(17-24\,\log2)-w_{1}\frac{8}{3}+\right.\right.\nonumber \\
 &  & +\left.w_{1}^{2}\frac{4}{45}\left(599-744\,\log2\right)\right]\alpha_{2}^{2}\nonumber \\
 &  & \left.\left.+(14+w_{1}^{2}\frac{124}{45})\,\alpha_{4}\right\} +...\right)\label{eq:osngf}
\end{eqnarray}
the first term in the parenthesis is divergent in the UV limit $w_{1}\to0$.
This divergence, as will be seen in the next section, is canceled
by the subtraction $S_{CT}$. The other terms are finite in this limit.
Also, in this limit the result for the coefficients of $a^{2}$ and
$a^{4}$ coincide with the ones in \cite{Andreev:2007vn}.

\subsection{Condensates of dimension $6,\;8$ and $10$}

It is recalled that as in subsection \ref{sub:In-QCD-} a condensate
of dimension $n$ is by definition the coefficient of $a^{n}$ in
the expansion of $\frac{\alpha'}{L^{2}}S_{NG}[z](a)$ in powers of
$a$. The calculation of these condensates is done in the UV limit,
$z_{1}\to0$. This procedure is valid since, according to the analysis
in subsection \ref{sub:convergence proof}, the only coefficient that
diverge in this limit, is the one corresponding to the perimeter of
the loop, i.e. the coefficient of $a^{1}.$ Taking into account this
remark, the calculation of these condensates follows the same technique
as in the previous subsection except that $z_{1}=0$ is taken from
the start. Their computation is possible under the assumption $\alpha_{2}=0$,
i.e. no dimension two condensate. As an example the condensate of
dimension $6$ is considered. That condensate must be proportional
to $\alpha_{6}$. This follows from the dimensional arguments which
are considered in appendix \ref{sub:Admissible-monomials-in}. There
it is shown that for the warp factor of the form (\ref{eq:warp}),
$\alpha'S_{NG}/L^{2}$ should be dimensionless, thus the coefficient
of $a^{6}$ in this quantity should have dimension of length to the
$-6$. Next recalling that the dimension of $\alpha_{n}$ is length
to the $-n$, then the only way of getting such a dimension in terms
of positive %
\footnote{See \ref{sub:UV-universality}%
} powers of the $\alpha$'s is by means of $\alpha_{2}\alpha_{4}$
or $\alpha_{6}$, thus the assumption $\alpha_{2}=0$ leaves only
$\alpha_{6}$. In terms of the variables $t$ and $\psi$ the action
to be considered is therefore,
\begin{equation}
S_{NG}^{(6)}=\frac{L^{2}}{\alpha'}\int_{0}^{1}\frac{e^{2a^{6}\alpha_{6}\psi^{3}}t\sqrt{4+\frac{\left(1+w_{1}^{2}-t^{2}\right)\psi'(t)^{2}}{t^{2}\psi\left(t\right)}}}{2\psi\left(t\right)}dt\label{eq:s6}
\end{equation}
next an expansion in powers of $a^{2}$ of the solution is considered
as in (\ref{eq:solpsi}), replacing this in the equation of motion
determines the coefficients $\psi_{i}(t)$, giving,
\begin{equation}
\psi^{(6)}(t)=t^{2}+a^{6}\frac{\alpha_{6}}{10}\left(24t-12t^{2}-6t^{4}-4t^{6}-24\,\log(1+t)\right)\label{eq:sol6}
\end{equation}
replacing in (\ref{eq:s6}) gives the following contribution proportional
to $a^{6}$,
\[
\frac{\alpha'}{L^{2}}S_{NG}|_{a^{6}}=\frac{3}{5}\alpha_{6}a^{6}
\]

For the case of the dimension $8$ condensate, for dimensional reasons,
only $\alpha_{4}$ and $\alpha_{8}$ are relevant. The action to be
considered thus involves only these two coefficients in the warp factor.
The solution can be obtained as a power series in $a$ up to order
$a^{8}$, having an expression considerably more lengthy than (\ref{eq:sol6}),
which is given in appendix C. The contribution proportional to $a^{8}$
to the NG action is given by,
\[
\frac{\alpha'}{L^{2}}S_{NG}|_{a^{8}}=-\frac{11}{5670}[(-2111+3360\,\log2)\alpha_{4}^{2}-270\alpha_{8}]a^{8}
\]

For the case of the dimension $10$ condensate, for dimensional reasons,
only $\alpha_{4},\,\alpha_{6}$ and $\alpha_{10}$ are relevant. The
action to be considered thus involves only these coefficients in the
warp factor. The solution can be obtained as a power series in $a$
up to order $a^{10}$, having an expression considerably more lengthy
than (\ref{eq:sol6}), which is given in appendix C. The contribution
proportional to $a^{10}$ to the NG action is given by,
\[
\frac{\alpha'}{L^{2}}S_{NG}|_{a^{10}}=\frac{13}{4725}[(2999-5040\,\log2)\alpha_{4}\alpha_{6}+175\alpha_{10}]a^{10}
\]

\section{The Gluon condensate, UV universality \label{sec:The-Gluon-condensate,}}

\subsection{The computation of the subtraction}

The subtracted NG action is,
\[
S_{NG}^{sub}=S_{NG}-S_{CT}
\]
where,
\begin{eqnarray}
S_{CT} & = & \frac{\sqrt{a^{2}+z_{1}^{2}}}{\alpha'}\int_{z_{1}}^{z_{m}}dz\, e^{2A(z)}\nonumber \\
 & = & L^{2}\frac{\sqrt{a^{2}+z_{1}^{2}}}{\alpha'}\int_{z_{1}}^{z_{m}}dz\,\frac{e^{2\sum_{n=1}\alpha_{n}z^{n}}}{z^{2}}\label{eq:ct}
\end{eqnarray}
and $z_{m}$ denotes the minimum of $e^{2A(z)}$. Here the computation
is done for the case where only $\alpha_{2}$ and $\alpha_{4}$ are
different from $0$. In this case,
\begin{equation}
z_{m}=\frac{1}{2}\sqrt{\frac{\sqrt{\alpha_{2}^{2}+4\alpha_{4}}}{\alpha_{4}}-\frac{\text{\ensuremath{\alpha_{2}}}}{\text{\ensuremath{\alpha_{4}}}}}\label{eq:min}
\end{equation}
Since the integrand in (\ref{eq:ct}) is well behaved in the integration
region then  the exponential in the integrand can be expanded before
doing the integral %
\footnote{For the case considered, $\alpha_{n}=\delta_{n4}\alpha_{4}$ , the
integral in (\ref{eq:ct}) can be explicitly calculated as,
\[
\int_{z_{1}}^{z_{m}}dz\,\frac{e^{2\alpha_{4}z^{4}}}{z^{2}}=\frac{1}{4}\left(\frac{E_{\frac{5}{4}}(-2\, z_{1}^{4}\alpha_{4})}{z_{1}}-\frac{E_{\frac{5}{4}}(-2\, z_{m}^{4}\alpha_{4})}{z_{m}}\right)
\]
\[
=\frac{1}{z_{1}}+\left(-\frac{E_{\frac{5}{4}}(-2\, z_{m}^{4}\,\alpha_{4})}{4z_{1}}+\frac{\Gamma(-\frac{1}{4})\left(-\alpha_{4}\right){}^{1/4}}{2\ 2^{3/4}}\right)
\]
\[
\cong-\frac{2\alpha_{4}z_{1}^{3}}{3}-\frac{2}{7}\alpha_{4\,}^{2}z_{1}^{7}+O(z_{1}^{8})
\]
where $E_{\nu}(z)$ denote the exponential integral and the last approximate
equality is an expansion in powers of $z_{1}$. From this expression
it is clear that the coefficients of positive powers of $z_{1}$ are
the same as the ones obtained expanding the integrand in (\ref{eq:ct}).%
} Proceeding in this way leads to,
\begin{eqnarray}
\frac{\alpha'}{L^{2}}S_{CT} & = & \sqrt{a^{2}+z_{1}^{2}}\int_{z_{1}}^{z_{m}}dz\,\frac{e^{2\alpha_{2}z^{2}+2\alpha_{4}z^{4}}}{z^{2}}\nonumber \\
 & = & \sqrt{a^{2}+z_{1}^{2}}\int_{z_{1}}^{z_{m}}dz\left[\frac{1}{z^{2}}+2\text{\ensuremath{\alpha_{2}}}\right.\nonumber \\
 &  & \left.+\left(2\alpha_{2}^{2}+2\text{\ensuremath{\alpha_{4}}}\right)z^{2}+\left(\frac{4\alpha_{2}^{3}}{3}+4\text{\ensuremath{\alpha_{2}}}\text{\ensuremath{\alpha_{4}}}\right)z^{4}+\cdots\right]\nonumber \\
 & = & \sqrt{a^{2}+z_{1}^{2}}\left[-\frac{1}{z}+2\text{\ensuremath{\alpha_{2}}}z+\frac{2}{3}\left(\text{\ensuremath{\alpha_{2}^{2}}}+\text{\ensuremath{\alpha_{4}}}\right)z^{3}\right.\nonumber \\
 &  & \left.+\frac{4}{15}\text{\ensuremath{\alpha_{2}}}\left(\text{\ensuremath{\alpha_{2}^{2}}}+3\text{\ensuremath{\alpha_{4}}}\right)z^{5}+\cdots\right]_{z_{1}}^{z_{m}}\label{eq:ct-2}
\end{eqnarray}
it is to be noted that the $-1/z$ appearing in the last equality,
when evaluated at $z=z_{1}$ and multiplied by the factor $\sqrt{a^{2}+z_{1}^{2}}$,
cancels the divergent term, when $z_{1}\to0$, appearing in the on-shell
NG action in (\ref{eq:osngf}).

\subsection{The subtracted on-shell NG action }

$S_{NG}^{sub}$ is defined in (\ref{eq:nambu-goto substraida}). Using
(\ref{eq:osngf}) , (\ref{eq:ct-2}) and keeping up to terms of order
$z_{1}^{2}$, leads to,
\begin{eqnarray}
\frac{\alpha'}{L^{2}}S_{NG}^{sub} & = & -1+a\left(\frac{1}{z_{m}}-2z_{m}\text{\ensuremath{\alpha_{2}}}\right)+a^{2}\left(\text{\ensuremath{\alpha_{2}}}\frac{10}{3}\right)\nonumber \\
 &  & +a^{4}(\ensuremath{\alpha_{2}^{2}}\frac{14}{9}\left(17-24\log2\right)+\alpha_{4}\frac{14}{9})\nonumber \\
 &  & -\frac{8}{3}a^{3}z_{1}\text{\ensuremath{\alpha_{2}^{2}}}+z_{1}^{2}\left[\alpha\left(\frac{13}{3}-\log16\right)\right.\nonumber \\
 &  & \left.+\alpha_{2}^{2}\frac{2}{45}\left(1198-1488\,\log2\right)-\frac{124}{45}\alpha_{4}\right]\label{eq:subng}
\end{eqnarray}
the first two lines corresponds to the terms that survive in the UV
limit $z_{1}\to0$. The IR scale $z_{m}$ should be replaced by its
expression (\ref{eq:min}). In this respect it is worth noticing that
the contribution of that scale is proportional to $a$, therefore
a change in that scale only changes the coefficient of the perimeter.
For $\alpha_{4}=0$ the results for the coefficients of $a^{2}$ and
$a^{4}$ coincide in the UV limit with the ones in \cite{Andreev:2007vn}. 

Next the expression of $S_{NG}^{sub}$ as a power series in $a^{2}$
in the UV limit $z_{1}\to0$, is given for the case $\alpha_{2}=0$,
\begin{eqnarray}
\frac{\alpha'}{L^{2}}S_{NG}^{sub}|_{z_{1=0}} & = & -1+\frac{1}{z_{m}}a+\alpha_{4}\frac{14}{9}a^{4}+\frac{3}{5}\alpha_{6}a^{6}\nonumber \\
 &  & -\frac{11}{5670}\left[(-2111+3360\,\log2)\alpha_{4}^{2}\right.\nonumber \\
 &  & \left.-270\alpha_{8}\right]a^{8}+\frac{13}{4725}\left[(2999\right.\nonumber \\
 &  & \left.-5040\,\log2)\alpha_{4}\alpha_{6}+175\alpha_{10}\right]a^{10}\label{eq:sub-ng-z1=00003D0}
\end{eqnarray}

Due to the proof in \ref{sub:UV-universality} these results are unchanged
by considering additional $\alpha$'s in the expression (\ref{eq:fz}).
It means that the coefficients of $a^{n}$ in this expression are
exact, the inclusion of additional terms in the expansion of the warp
exponent do not change their value. This is a strictly UV result,
it is only valid in the limit $z_{1}\to0$. It shows that if the expectation
value of Wilson loops are related to minimal areas in the dual theory,
as assumed, then the expectation values of gauge invariant operators
in QCD can be used to systematically build the QCD dual background.
In particular, since there is no dimension two gauge invariant operator
in pure QCD then the coefficient of $a^{2}$ should be zero  %
\footnote{See footnote {[}30{]} .%
}. Thus, under these conditions, this absence implies $\alpha_{2}=0$.
The case of the coefficient of $a^{4}$ is different since in QCD
there is a gauge invariant quantity of dimension $4$, which is the
expectation value of $\langle F_{\mu\nu}F^{\nu\mu}\rangle$. This
coefficient is related to the gluon condensate, and its value fixes
the value of $\alpha_{4}$. This procedure can be continued for higher
order terms in the expansion. Higher dimensional condensates fix the
values of higher index $\alpha_{i}$ coefficients, once the ones with
lower indices are known. This is clearly exemplified by expression
(\ref{eq:sub-ng-z1=00003D0}).

\subsection{Computation of the gluon condensates}

For the soft wall case $\alpha_{i}=\delta_{i,2}\alpha_{2}$ the results
are the same as in \cite{Andreev:2007vn}.

\noindent For the $z^{4}$case $\alpha_{i}=\delta_{i4}\alpha_{4}$.
Eq. (\ref{eq:subng}) shows that in this case the coefficient of $a^{4}$
in $S_{NG}^{sub}$ is,

\noindent 
\[
\frac{14}{9}\frac{L^{2}}{\alpha'}\alpha_{4}\;.
\]
Using the expression of $G_{2}$ in terms of this coefficient appearing
in \cite{Andreev:2007vn} leads to the following expression for $\alpha_{4}$,
\begin{equation}
\frac{L^{2}}{\alpha'}\frac{14}{9}\alpha_{4}=\frac{\pi^{4}}{36}G_{2}\;,\label{eq:subz4}
\end{equation}
which according to the value of $G_{2}=0.028\, GeV^{4}$ in \cite{Ilgenfritz:2012nca}
leads to,
\[
\frac{L^{2}}{\alpha'}\alpha_{4}=0.0049\, GeV^{4}\;.
\]

\noindent In the appendix it is shown how an additional relation between
$\frac{L^{2}}{\alpha'}$ and $\alpha_{4}$ can be obtained by means
of computing the string tension for the linear potential between static
quarks %
\footnote{This argument is based on the important fact that the coefficient
in front of the NG action, i.e. $\frac{L^{2}}{\alpha'}$ , is independent
of the loop's shape. In particular it is the same for the circular
and for the rectangular loops. %
} in the case $\alpha_{i}=\delta_{i4}\alpha_{4}$. This relation is,
\[
\sigma=\frac{\alpha'}{L^{2}}2\sqrt{e}\sqrt{\text{\ensuremath{\alpha_{4}}}}
\]
where $\sigma$ is the string tension mentioned above. Taking for
it the slope in the linear term of the Cornell potential\cite{Cornell:75},
i.e. $\sigma=0.186\, GeV^{2}$ leads to, 
\[
\frac{L^{2}}{\alpha'}=1.155\;\;,\alpha_{4}=0.00424\,\text{GeV}^{4}
\]
For the case $\alpha_{i}=\delta_{i6}\alpha_{6}$ the coefficient of
$a^{6}$ in $S_{NG}^{sub}$ is,
\begin{equation}
\frac{L^{2}}{\alpha'}\frac{3}{5}\alpha_{6}\label{eq:a6}
\end{equation}

\noindent In \cite{Shifman1979385} there is an estimation of the
following dimension $6$ expectation value given by,
\[
\langle g^{3}f_{abc}F_{\alpha\beta}^{a}F_{\beta\delta}^{b}F_{\delta\alpha}^{c}\rangle\backsimeq0.045\text{GeV}^{6}
\]
in addition in \cite{Shifman:1980ui} a relation between this expectation
value and the coefficient of $a^{6}$ in the expansion of a circular
loop on its radius $a$ is derived,
\[
\left.\langle W\left(C\right)\rangle\right\rfloor _{a^{6}}=\frac{\pi^{2}}{192N_{c}}\langle g^{3}f_{abc}F_{\alpha\beta}^{a}F_{\beta\delta}^{b}F_{\delta\alpha}^{c}\rangle
\]
this should be equal to the coefficient (\ref{eq:a6}). Furthermore
as shown in appendix \ref{sub:Linear-potential-between} the string
tension for this case is given by,
\[
\sigma=\frac{\alpha'}{L^{2}}(6\alpha_{6}e)^{1/3}
\]
thus the following two relations involving $\frac{L^{2}}{\alpha'}$
and $\alpha_{6}$ are obtained,
\begin{eqnarray*}
\frac{L^{2}}{\alpha'}\frac{3}{5}\alpha_{6} & = & \frac{\pi^{2}}{192N_{c}}0.045\text{GeV}^{6}\\
0.186\, GeV^{2} & = & \frac{\alpha'}{L^{2}}(6\alpha_{6}e)^{1/3}
\end{eqnarray*}
 where as before the string tension is taken to be $\sigma=0.186\, GeV^{2}$.
Solving these equations for $\frac{L^{2}}{\alpha'}$ and $\alpha_{6}$,
leads to,
\[
\frac{L^{2}}{\alpha'}=1.336\;\;,\alpha_{6}=0.00094\,\text{GeV}^{6}
\]

\section{Concluding remarks}

In this work the calculation of the minimal area bounded by a circular
loop lying at a certain value $z_{1}$ of the radial coordinate $z$
has been considered. This surface is embedded in a 5-dimensional space
with a global metric which in conformal coordinates depends only on
the warp factor $e^{2A(z)}$. The connection of this calculation with
QCD observables is shown in the following scheme,
\[
Global\; metric\overset{NG}{\longleftrightarrow}Min.\; area\overset{a-exp.}{\longleftrightarrow}QCD\; condensates
\]
where $a-exp.$ goes for the expansion of the minimal area in powers
of the loop radius $a$. The continuation of this scheme to the left
would require the knowledge of a gravity-string theory from which
the warp factor could be obtained. In this respect it is worth remarking
that if such a theory would include a dilaton field then the warp
factor $e^{2A(z)}$ considered in this work corresponds to the string
frame warp factor \cite{Gursoy:2007er}. The arrows in the above scheme
go in both directions, trying to indicate that these connections could
be employed in both ways. That is, knowledge of QCD condensates could
be employed to obtain warp factors as in (\ref{eq:subz4}) and, in
the other direction, details of a higher dimensional theory would
give information about QCD.

Regarding the connection between the minimal area and the condensates,
it is emphasized that an important ingredient for this connection
is the subtraction employed. This subtraction involves both UV and
IR divergences, the first already present in the AdS case are treated
as in \cite{Maldacena:1998im} and maintaining conformal invariance,
the second coming form the consideration of confining warp factors,
require an IR scale which is argued to be given naturally by the location
of the minimum of these warp factors. In this respect, it is important
to realize that the approximations employed are well suited for the
calculation of the first coefficients in the expansion in powers of
the radius $a$ for the \emph{subtracted} NG action. 

Finally it is noted that the techniques employed in this work are
not restricted to the particular family of warp factors (\ref{eq:warp}).
Any other choice that can be made convergent by the subtractions appearing
in section \ref{sec:The-substracted-on-shell} would work  %
\footnote{See Appendix \ref{sub:The-substraction-for} for an example.%
}. If this is not the case other subtractions should be considered.

This work was supported by DOE Contract No. DE-AC05-06OR23177 under
which JSA operates the Thomas Jefferson National Accelerator Facility,
and by the National Science Foundation (USA) through grants PHY-0855789
and PHY-1307413 (J.L.G.), and by CONICET (Argentina) PIP Nº 11220090101018
(R.C.T.). J.L.G. thanks the Instituto Balseiro and the Centro At\textbackslash{}'omico
Bariloche for hospitality and support during the early stages of this
project.

\section*{Appendices}

\appendix*

\setcounter{equation}{0}

\subsection{The subtraction for $Dp$-branes inspired warp factors\label{sub:The-substraction-for}}

As an example of other warp factors of interest , the following are
considered,
\begin{equation}
A(z)=-n\,\log\left(\frac{z}{L}\right)+f(z)\label{eq:warp-n}
\end{equation}
The case $n=1$ is the one already studied in \ref{sub:convergence proof}.
For backgrounds generated by a stack of $Dp$-branes, one often arrives
to metrics with $n\leq1$. These warp factors diverge in the UV, the
leading singularity is,

\begin{equation}
A\left(z\right)\sim-n\log\left(\frac{z}{L}\right)\Rightarrow A'\left(z\right)\sim-n\frac{L}{z}\label{eq:az0-1}
\end{equation}
the equation of motion near the boundary implies,
\[
-\underset{z\rightarrow0}{\lim}2\, r\, n\,\frac{1}{z}r'\left(1+r'^{2}\right)+\underset{z\rightarrow0}{\lim}rr''-\underset{z\rightarrow0}{\lim}\left(1+r'^{2}\right)=0
\]
which in a similar way as in \ref{sub:convergence proof} leads to
the following asymptotic behavior,
\[
\underset{z\to0}{\lim}r'\left(z\right)=0r'\left(z\right)=\frac{1}{a\left(1-2n\right)}z+...\left(z\ll1\right)
\]
Inserting this in $S_{NG}^{sub}$, shows that the leading behavior
of the integrand in the NG action is governed by,
\begin{eqnarray*}
\underset{z\to0}{\lim}e^{2A\left(z\right)}\left(r\left(z\right)\sqrt{1+r'\left(z\right)^{2}}-a\right) & =\\
\underset{z\to0}{\lim}\frac{1}{z^{2n}}\left(a\left(1+\frac{1}{2}c_{0}^{2}z^{2}+\dots\right)-a\right) & =\\
a\underset{z\to0}{\lim}\frac{1}{z^{2n}}\left(\frac{1}{2}c_{0}^{2}z^{2}+\dots\right) & = & 0
\end{eqnarray*}
This implies that for $n>1$ the regularization procedure does not
work since the expression diverges. However $n\leq1$ corresponds
to the metrics obtained from top-bottom approaches with stacks of
$Dp$-branes. A concrete example can be found in \cite{Kopnin:2011si},
where the area of the circular loop is found to be,

\begin{equation}
S_{NG}|_{Dp}=\frac{1}{2\pi\alpha'}\int_{0}^{a}\left(\frac{5-p}{2}\frac{1}{z}\right)^{\frac{7-p}{5-p}}r\sqrt{1+z'^{2}}dr
\end{equation}
The regularization procedure ensures the convergence of the subtracted
area except for the cases $p=4$ and $p=5$.

\subsection{Admissible monomials in the expansion of the NG action solution\label{sub:Admissible-monomials-in}}

The NG action times $\alpha'$ is an area and therefore has dimension
of length squared. Making explicit the first term in (\ref{eq:warp-n})
it is written as follows,
\[
\alpha'S_{NG}=L^{2n}\int_{0}^{a}\frac{e^{2\sum_{k=1}\alpha_{k}z^{k}}}{z^{2n}}r\sqrt{1+z'^{2}}dr
\]
therefore the integral in this last equation should have dimensions
of length to the power $2-2n$. In particular for the case $n=1$
(deformation of AdS), it should be dimensionless. This integral depends
on $a\;,z_{1}$,which have dimension of length, and the $\alpha$'s.
In this respect it is useful to note that the $\alpha_{k}$ has dimensions
of length to the power $-k$. Any monomial contributing to $\frac{\alpha'}{L^{2n}}S_{NG}$
of the general form,
\[
a^{j}z_{1}^{m}\alpha_{k}^{l}
\]
will have vanishing coefficient unless, 
\[
j+m-k\cdot l=2-2n
\]
The same general conclusions are valid for $\frac{\alpha'}{L^{2n}}S_{CT}$.

\subsection{Solutions needed to obtain the condensates of dimensions $8$ and
$10$}

\begin{widetext}

\begin{eqnarray*}
\psi^{(8)}(t) & = & t^{2}-\frac{2}{3}a^{4}\left(t\left(-4+2t+t^{3}\right)+4\log(1+t)\right)\alpha_{4}+\frac{1}{945}a^{8}\left\{ -2\left[560\pi^{2}+t[-3824\right.\right.\\
 &  & +t\,\left(5272+t\,\left(1260-304t-156\, t^{3}+135t^{5}\right)\right)]+13440\, t\:\log2-3360\:\log^{2}2\\
 &  & +6720\:\text{log}(1-t]\log(\frac{2}{1+t})-8\:\log(1+t)\left(-478+105t\left(6+6t+t^{3}\right)\right.\\
 &  & \left.\left.+210\;\log(1+t)\right)-6720\:\mathrm{Li_{2}}(1+t)\right]\alpha-90\left(t\left(-24+12t+6t^{3}+4t^{5}+3t^{7}\right)\right.\left.\left.+24\log(1+t)\right)\alpha_{8}\right\} 
\end{eqnarray*}
\begin{eqnarray*}
\psi^{(10)}(t) & = & t^{2}-\frac{2}{3}a^{4}\left(t\left(-4+2t+t^{3}\right)+4\log(1+t)\right)\alpha_{4}-\frac{2}{945}a^{8}\left[560\pi^{2}+t\left(-3824+t\right.\right.\\
 &  & \left(\left(5272+t\left(1260-304t-156t^{3}+135t^{5}\right)\right)\right)+13440\, t\:\log2-3360\log^{2}2+6720\log(1-t)\;\log(\frac{2}{1+t})\\
 &  & -8\log(1+t)\left(-478+105t\left(6+6t+t^{3}\right)+210\log(1+t)\right)-6720\:\mathrm{Li_{2}}(\frac{1+t}{2})\alpha_{4}^{2}\\
 &  & -\frac{1}{5}a^{6}\left(t\left(-12+6t+3t^{3}+2t^{5}\right)+12\log(1+t)\right)\alpha_{6}+a^{10}\left(\frac{1}{945}\left(-2520\pi^{2}+15120\,(\log2)^{2}\right.\right.\\
 &  & -t(-22128+t\left(26184+t\left(2604+t\left(240+t\left(2772-1772\, t-321t^{3}+420\, t^{5}\right)\right)\right)\right)\\
 &  & +60480\log2)-30240\;\log(1-t)\,\log(\frac{2}{1+t})+24\log(1+t)\left(-922+21t\left(48+42t+15t^{3}+2t^{5}\right)\right.\\
 &  & \left.+252\log(1+t))+30240\,\mathrm{Li_{2}}(1+t)\right)\alpha_{4}\alpha_{6}-\frac{1}{54}\left(t\left(-120+60t+30t^{3}+20t^{5}+15t^{7}+12t^{9}\right)\right.\\
 &  & \left.+120\log(1+t))\alpha_{10}\right),
\end{eqnarray*}

\end{widetext}

where $\mathrm{Li_{2}}$ denotes the dilogarithm function.

\subsection{Linear potential between static quarks \label{sub:Linear-potential-between}}

The string tension is given by the value at its minimum of the function
\cite{kinar-sonnenschein:1998},
\[
f(z)=\frac{\alpha'}{L^{2}}e^{2A(z)}
\]
for the case $\alpha_{n}=\alpha_{4}\delta_{n4}$ it is given by,
\[
f(z)=\frac{\alpha'}{L^{2}}e^{2(-\log z+\alpha_{4}z^{4})}
\]
its minimum and the corresponding string tension are,
\[
z_{0}=\frac{1}{\sqrt{2}\text{\ensuremath{\alpha_{4}^{1/4}}}}\;\;,\sigma=f(z_{0})=\frac{\alpha'}{L^{2}}2\sqrt{e}\sqrt{\text{\ensuremath{\alpha_{4}}}}
\]
which has the right units since $\alpha_{4}$ has units of length
to the minus $4$, thus $\sqrt{\alpha_{4}}$ has units of energy squared
as corresponds to a string tension. 

For the case $\alpha_{n}=\alpha_{6}\delta_{n6}$ the minimum and string
tension are given by,
\[
z_{0}=\frac{1}{(6\alpha_{6}){}^{1/6}}\;\;\;,f(z_{0})=\frac{\alpha'}{L^{2}}(6\alpha_{6}e)^{1/3}
\]

\bibliographystyle{unsrt}
\addcontentsline{toc}{section}{\refname}\bibliography{../Bibliography}

\end{document}